\documentclass{ifacmtg}
\usepackage{amsmath}
\usepackage[dvips]{graphics}

\usepackage{epsfig}
\usepackage[dcucite]{harvard}
\newtheorem{definition}{Definition}[section]

\newtheorem{theorem}[definition]{Theorem}
\newtheorem{proposition}[definition]{Proposition}

\newtheorem{example}[definition]{Example}

 1   
\font\ddpp=msbm10  scaled \magstep 1  
\font\ddppp=msbm8  scaled \magstep 1  
\def\bull{\ \ \ \vrule height 1.5ex width.8ex depth.3ex \medskip}

\def\QED{\hskip0.1em\hfill\null\ \null\nobreak\hfill
\kern3pt\lower1.8pt\vbox{\hrule\hbox   {\vrule\kern1pt\vbox{\kern1.7pt
\hbox{$\scriptstyle   QED$}\kern0.2pt}\kern1pt\vrule}\hrule}}

\def\R{\hbox{\ddppp R}}               
\def\S{\hbox{\ddpp S}}  
\def\ene{\hbox{\ddppp N}}    

\def\hfl#1#2{\smash{\mathop{\hbox to 12 mm{\rightarrowfill}}
\limits^{\scriptstyle#1}_{\scriptstyle#2}}}

\begin{document}
\runauthor{M. de Le\'on, D. Mart{\'\i}n de Diego, A. Santamar{\'\i}a Merino}
\begin{frontmatter}
\title{Geometric numerical integration of nonholonomic systems and optimal control problems}
\author[P]{M. de Le\'on}
\author[P]{D. Mart{\'\i}n de Diego}
\author[P]{A. Santamar{\'\i}a Merino\thanksref{Someone}}
\address[P]{Instituto de Matem\'aticas y F\'\i sica Fundamental,
CSIC,
Serrano 123, 
28006 Madrid, Spain}
\thanks[Someone]{This work has been  supported by grant BFM2001-2272. A. Santamar{\'\i}a Merino wishes to thank the Programa de formaci\'on de Investigadores of the Departamento de Educaci\'on, Universidades e Investigaci\'on of the Basque Government (Spain) for financial support.
}
\begin{abstract}
A geometric derivation of  numerical integrators for nonholonomic systems  and optimal control problems is obtained.
It is based in the classical technique of generating functions adapted to the special features of nonholonomic systems and optimal control problems. 
\end{abstract}
\end{frontmatter}

\section{Introduction}

Standard methods for simulating the motion of a dynamical system usually ignore many of  the geometric features of this system (simplecticity, conservation laws, symmetries...).
 However, new methods have been recently developed, called geometric integrators, which are concerned with  some of the extra features of  geometric nature of the dynamical system (see [HaLuWa:02]). 

In the first part of the paper, we propose  a class of geometric   integrators for nonholonomic systems  [Leomar:96D,NeiFuf:72] based on a discretization of the Lagrangian function (in a more precise sense, we discretize the action function) and a coherent discretization of the constraint forces (see [LeMaSa:02]).
These  equations will be  conceptually equivalent to the proposed for systems with external forces (see [MarWes:01]).
Finally,   second part corcerns  with the construction of symplectic integrators for optimal control theory by using generating functions of the second kind. 

\section{Nonholonomic systems}
\subsection{Geometrical formulation of non\-ho\-lo\-no\-mic systems}

Let $Q$ be a $n$-dimensional differentiable manifold, with local coordinates
$(q^i)$ and  tangent bundle $TQ$, with induced coordinates $(q^i, \dot q^i)$.
 Consider a Lagrangian system, with Lagrangian $L: TQ \rightarrow \R$, subject
to nonholonomic constraints, defined by a submanifold $D$ of the velocity phase
space $TQ$. We will assume that $\dim D = 2n - m$ and that $D$ is locally 
described by the vanishing of $m$ independent functions $\phi^a$ (the
``constraint functions"), satisfying the rank condition $\displaystyle{\hbox{rank }\left(\frac{\partial \phi^a}{\partial \dot{q}^i}\right)=m}$.
In the sequel, we will follow a Hamiltonian point of view. The canonical
coordinates on $T^*Q$ (the cotangent bundle of $Q$)  are denoted by $(q^i, p_i)$. Assume, for simplicity, that the Lagrangian $L$ is hyperregular, that is,  the 
Legendre transformation $Leg: TQ \rightarrow T^*Q, (q^i, \dot{q}^i) \mapsto
(q^i, p_i = \partial L/\partial \dot{q}^i)$, is a global diffeomorphism.
The constraint functions on $T^*Q$ become $\Psi^a=\phi^a\circ Leg^{-1}$, i.e.
$\displaystyle{\Psi^a(q^i, p_i)=\phi^a(q^i, \frac{\partial H}{\partial p_i})\,,}
$
where the Hamiltonian $H: T^*Q\rightarrow \R$ is defined by  
$H=E_L\circ Leg^{-1}$. Here, $E_L$ denotes  the energy of the system, locally defined by $E_L=\dot{q}^i\frac{\partial L}{\partial \dot{q}^i}-L$. 
Since locally 
$Leg^{-1}(q^i, p_i)=(q^i, \displaystyle{\frac{\partial H}{\partial p_i}})$, then
$
H=p_i \dot{q}^i-L(q^i, \dot{q}^i)\; ,
$
where $\dot{q}^i$ is expressed in terms of $q^i$ and $p_i$ by using $Leg^{-1}$.

The equations of motion for the nonholonomic system on
$T^*Q$ can now be written as follows (see [CaLeMa:99,Marl:95] and references therein)
\begin{equation}\label{hnh}
\left\{
\begin{array}{rcl}
\dot q^i&=&\displaystyle{\frac{\partial H}{\partial p_i}}\\
\vphantom{\huge A}\dot p_i&=&\displaystyle{-\frac{\partial H}{\partial q^i}-{\lambda}_a \frac{\partial
\Psi^a}{\partial p_j}{\cal H}_{ji}}\, ,
\end{array}\right.
\end{equation}
together with the constraint equations 
$\Psi^a(q,p)=0$, 
where ${\cal H}_{ij}$ are the components of the inverse of the matrix 
$({\cal H}^{ij})=(\partial^2 H/ \partial p_i\partial p_j)$. Note that
\[
(\frac{\partial \Psi^a}{\partial p_j}{\cal H}_{ji})(q,p) = 
(\frac{\partial \phi^a}{\partial \dot{q}^i} \circ Leg^{-1})(q,p).
\]
 Let $M$
denote the image of the constraint submanifold $D$ under the Legendre
transformation, and let $F$ be the distribution 
on $T^*Q$ along $M$, whose annihilator is given by
$F^o = Leg_*( \tilde{F}^o))$. Here, $\tilde{F}^o$ represents the constraint forces subbundle, locally defined by
\[
\tilde{F}^o=\hbox{span}\{\mu^a=\frac{\partial \phi^a}{\partial \dot{q}^i}\, dq^i\}
\]
The Hamiltonian equations of motion of the nonholonomic system can be then
rewritten in intrinsic form as
\begin{equation}\label{a1}
\begin{array}{rcl}
(i_X\omega_Q-dH)_{|M}&\in& F^{o}\\
 X_{|M} &\in& TM \,,
\end{array}
\end{equation}
where $\omega_Q=-d\theta_Q=dq^i\wedge dp_i$ (with $\theta_Q=p_i\, dq^i$)   is the canonical symplectic form on $T^*Q$.
Suppose in addition that the following {\em compatibility condition} $F^{\perp}\cap TM=\{0\}$ holds, where $``\perp"$ denotes the symplectic orthogonal with respect to $\omega_Q$. 
Observe that, locally, this condition means that the matrix
$\displaystyle{
({C}^{ab})=
\left(\frac{\partial \Psi^a}{\partial p_i}{\cal H}_{ij}\frac{\partial
\Psi^b}{\partial p_j}\right)
}$
is regular.  
The compatibility condition is not too restrictive, since it is trivially verified by the 
usual systems of mechanical type (i.e.\ with a Lagrangian of the form kinetic
minus potential energy), where the ${\cal H}_{ij}$ represent the components of
a positive definite Riemannian metric. 
The compatibility condition  guarantees, in particular, the existence of a unique
solution of the constrained  equations of motion (\ref{a1}) which, henceforth,
will be denoted by $X_{H,M}$ on the Hamiltonian side and $Leg^{-1}_*(X_{H,M})=\xi_{L,D}$ on the Lagrangian side.

Moreover, if we denote by $X_H$ the Hamiltonian function of $H$, i.e.,
$i_{X_H}\omega_Q=dH$ then, using the constraint functions, we may explicitely determine the
Lagrange multipliers $\lambda_a$ as
$\lambda_a=  -{\cal C}_{ab} X_H(\Psi^b)\; .
$
Next, writing the 1-form  
$
\Lambda=-{\cal C}_{ab} X_H(\Psi^b)\frac{\partial
\Psi^a}{\partial p_j}{\cal H}_{ji}dq^i 
$
then, the nonholonomic equations are equivalently rewritten as
\begin{equation}\label{hnh1}
\left\{
\begin{array}{rcl}
\dot q^i&=&\displaystyle{\frac{\partial H}{\partial p_i}}\; ,\\
\dot p_i&=&\displaystyle{-\frac{\partial H}{\partial q^i}-\Lambda_i}\, ,
\end{array}\right.
\end{equation}
for initial conditions $(q_0, p_0)\in M$ and $\Lambda=\Lambda_i\, dq^i$. We also denote by $\tilde{\Lambda}={Leg}^*({\Lambda})$ the 1-form on $TQ$ wich represents the constraint force   
once the Lagrange multipliers have been determined.

Now, consider the flow 
$F_t: M\rightarrow M$, $t\in I\subseteq \R$  of the vector field $X_{H, M}$, solution of the nonholonomic problem.
Since (\ref{hnh1}) is geometrically rewritten as
\[
i_{X_{H, M}}\omega_Q=dH+\Lambda\; ,
 \]
then
\[
L_{X_{H, M}}\theta_Q=d(i_{X_{H,M}}\theta_Q-H)-\Lambda\; ,
\]
or, equivalently,
$
L_{X_{H, M}}\theta_Q=d(L\circ Leg^{-1})-\Lambda\; .
$
Therefore, integrating
\begin{equation}\label{ei}
\hskip-0.4cm F_h^*\theta_Q-\theta_Q=d\left(\int^h_0 L\circ \tilde{F}_t \,dt\right)-\int^h_0 F_t^*\Lambda\; ,
\end{equation}
where $\tilde{F}_t$ is the flow of the vector field $\xi_{L, D}$.

\subsection{``Generating  functions" and nonholonomic mechanics}\label{subsection}

In what follows, we will follow similar arguments for the construction of generating functions for symplectic or canonical maps [Arn:78]. However, because of equation (\ref{ei}), we have that the nonholonomic flow is not a canonical transformation; i.e.,
\begin{equation}\label{ei1}
F_h^*\omega_Q-\omega_Q=d\left(\int^h_0 F_t^*\Lambda\right)\; .
\end{equation}
 This description will allow us to construct a new family of nonholonomic integrators for equations (\ref{aqq}). 
Denote by $\pi_i: T^*Q\times T^*Q\rightarrow T^*Q$, $i=1,2$, the canonical projections. Consider the following forms
\begin{eqnarray*}
{\Theta}&=&\pi_2^*\theta_Q-\pi_1^* \theta_Q\; ,\\
{\Omega}&=&\pi_2^*\omega_Q-\pi_1^* \omega_Q=-d{\Theta}\; .
\end{eqnarray*}
Denote  by $i_{F_h}: \hbox{Graph}(F_h)\hookrightarrow T^*Q\times T^*Q$ the inclusion map and observe  that
$\hbox{Graph}(F_h)\subset M\times M$.
Then, from (\ref{ei})   
$
i_{F_h}^*{\Theta}$ is equal to 
\[
({\pi_1}_{|\hbox{\tiny  Graph}(F_h)})^*\left[ d\left(\int^h_0 L\circ \tilde{F}_t \,dt\right)-\int^h_0 F_t^*{\Lambda}\right].
\]


Let  $(q_0, p_0, q_1, p_1)$  be coordinates in $T^*Q\times T^*Q$ in a neighborhood of some point in  $\hbox{Graph}(F_h)$. If   $(q_0, p_0, q_1, p_1)\in \hbox{Graph}(F_h)$ then  
$\Psi^a(q_0, p_0)=0$ and $\Psi^a(q_1,  p_1)=0$.
Moreover,  along $\hbox{Graph}(F_h)$, $q_1=q_1(q_0,p_0)$, $p_1=p_1(q_0,p_0)$ and 
\begin{eqnarray}\label{fr}
p_1\,dq_1-p_0dq_0&=&d \left(\int^h_0 L(q(t), \dot{q}(t)) \,dt\right)\nonumber
\\&&-\int^h_0 \widetilde{\Lambda}(q(t),\dot{q}(t)), 
\end{eqnarray}
where $(q(t), \dot{q}(t))=\tilde{F}_t(q_0, \dot{q}_0)$ with $Leg(q_0, \dot{q}_0)=(q_0, p_0)$. Here, $\tilde{F}_t$ denotes the flow of $\xi_{L,D}$. Equation (\ref{fr}) is satisfied along $\hbox{Graph}(F_h)$.

Assume that, in a neighborhood of some point $x\in \hbox{Graph}(F_h)$, we can change this system of coordinates to a  new  coordinates 
$(q_0, q_1)$. 
Denote by 
\[
S^h(q_0, q_1)=\int^h_0 L(q(t), \dot{q}(t))\, dt\, ,
\]
where $q(t)$ is a solution curve  of the nonholonomic problem with $q(0)=q$ and $q(h)=q_1$ and an adequate extension of $S^h$. It is easy to show that this  solution always exists for adequate values of $q_0$ and $q_1$. 

Thus,  we deduce  
 that
\begin{equation}\label{aq1}
\left\{
\begin{array}{l}
 \displaystyle{p_0= -\frac{\partial S^h}{\partial q_0}}+\int^h_0 \widetilde{\Lambda}(q(t),\dot{q}(t))\frac{\partial q}{\partial q_0}\; ,\\
\displaystyle{p_1= \frac{\partial S^h}{\partial q_1}}-\int^h_0 \widetilde{\Lambda}(q(t),\dot{q}(t))\frac{\partial q}{\partial q_1}\; ,\\
\end{array}
\right.
\end{equation}
where $(q_0, q_1)$  verifies the constraint functions $\varphi^a(q_0, q_1,h)=0$,  explicitely  defined by 
\begin{eqnarray}\label{qwe}
\hskip-0.5cm&&\varphi^a(q_0, q_1,h)=\nonumber\\
\hskip-0.5cm&& \Psi^a(q_0, -\frac{\partial S^h}{\partial q_0}(q_0, q_1)+\int^h_0 \widetilde{\Lambda}(q(t),\dot{q}(t))\frac{\partial q}{\partial q_0}),
\end{eqnarray}
where $q(t)$ is a solution of the nonholonomic problem with $q(0)=q_0$ and $q(h)=q_h$.

Next, we will show how the group composite law of the flow $F_h$, 
$\displaystyle{F_{Nh}=\underbrace{F_h\circ\ldots \circ F_h}_{N}}$, 
is expressed in terms of the corresponding ``generating functions" $S^h$. 
Moreover, the following Theorem will result in  a new construction of numerical integrators for nonholonomic mechanics when we change the ``generating function" and the constraint forces by appropriate approximations. 
\begin{theorem}\label{Th}
The function $S^{Nh}$, the ``generating function" for $F_{Nh}$, is  given  by
\[
S^{Nh}(q_0, q_{N})=\sum_{k=0}^{N-1}S^h(q_k, q_{k+1})\; ,
\]
where $q_k$, $1\leq k\leq N-1$, are  points verifying
\begin{eqnarray}\label{ase}
\hskip-0.7cm&& D_2 S^h(q_{k-1}, q_k)+D_1 S^h(q_k, q_{k+1})=\nonumber\\
\hskip-0.7cm&&\int^h_0\widetilde{\Lambda}(q(t), \dot{q}(t))\frac{\partial q}{\partial q_1}+\int^{2h}_h\widetilde{\Lambda}(q(t), \dot{q}(t))\frac{\partial q}{\partial q_0},
\end{eqnarray}
and
$q(t)$ is a solution curve  of the nonholonomic problem with $q(0)=q_{k-1}$ and $q(h)=q_{k}$ 
(respectively, $q(h)=q_k$ and $q(2h)=q_{k+1}$) for the first integral (resp., second integral) of the right-hand side.
\end{theorem}
{\bf Proof:}
It is suffices to prove the result for $N=2$; that is, 
\[
S^{2h}(q_0, q_2)=S^h(q_0, q_1)+S^h(q_1, q_2)\; ,
\]
where $q_1$ verifies condition (\ref{ase}).

Since
\begin{eqnarray*}
p_1\,dq_1-p_0\, dq_0&=&dS^h(q_0, q_1)-\int^h_0 \widetilde{\Lambda}(q(t),\dot{q}(t))\; ,\\
p_{2}\,dq_{2}-p_1\, dq_1&=&dS^h(q_1, q_2)-\int^{2h}_h \widetilde{\Lambda}(q(t),\dot{q}(t))\; ,
\end{eqnarray*}
then 
\begin{eqnarray*}
&&p_2\,dq_2-p_0\, dq_0=d\left(S^h(q_0, q_1)+S^h(q_1, q_2)\right)\\
&&-\int^h_0 \widetilde{\Lambda}(q(t),\dot{q}(t))-\int^{2h}_h \widetilde{\Lambda}(q(t),\dot{q}(t))\; .
\end{eqnarray*}
Since the variables $q_1$ do not appear on the left-hand side term, we obtain expression 
(\ref{ase}). Moreover, for this choice of $q_1$ then 
$
S^{2h}(q_0, q_2)=S^h(q_0, q_1)+S^h(q_1, q_2)
$
is a ``generating function of the first kind" of $F_{2h}$. \bull

Equations (\ref{ase}) determine an implicit system of difference equations which permit us to obtain $q_2$ from the initial data $q_0$ and $q_1$.

\vspace{-0.5cm}

\subsection{ Nonholonomic integrators}

In the sequel and,  for simplicity, assume that $Q$ is a vector space.
Since $S^h(q_0, q_1)=\int^h_0 L(q(t), \dot{q}(t))\, dt$, where $q(t)$ is a nonholonomic solution with $q(0)=q_0$ and $q(h)=q_1$,
we can obtain nonholonomic integrators by taking adequate approximations of the ``generating function" $S^h$ and the extra-term 
$\int^h_0\tilde{\Lambda}(q(t), \dot{q}(t))$.

Consider, for instance,  the approximation
\begin{equation}\label{lol}
\hskip-0.5cm S^h_{\alpha}(q_0, q_1)=hL((1-\alpha)q_0+\alpha q_1, \frac{q_1-q_0}{h})\; ,
\end{equation}
for some parameter $\alpha\in [0,1]$.  (In general, we will write $S^h_{\alpha}(q_0, q_1)\approx S^h(q_0, q_1)$.)

A natural approximation of the constraint forces adapted to our choice of approximation for $S^h$ are
\begin{eqnarray*}
\hskip-0.2cm&&\int^h_0\widetilde{\Lambda}(q(t), \dot{q}(t))\frac{\partial q}{\partial q_0}\\
\hskip-0.2cm&&\approx 
(1-\alpha) h \widetilde{\Lambda}((1-\alpha)q_0+\alpha q_1, \frac{q_1-q_0}{h})\; ,\\
\hskip-0.2cm&&\int^{h}_{0}\widetilde{\Lambda}(q(t), \dot{q}(t))\frac{\partial q}{\partial q_1}\approx 
\alpha h \widetilde{\Lambda}((1-\alpha)q_0+\alpha q_1, \frac{q_1-q_0}{h})\; .
\end{eqnarray*}

Consequently, we obtain the  following numerical method for nonholonomic systems 
\begin{eqnarray*}
&&D_2 S^h_{\alpha}(q_{k-1}, q_k)+D_1 S^h_{\alpha}(q_k, q_{k+1})=
\\
&&\alpha h \widetilde{\Lambda}((1-\alpha)q_{k-1}+\alpha q_k, \frac{q_k-q_{k-1}}{h})\\
&&+(1-\alpha) h \widetilde{\Lambda}((1-\alpha)q_k+\alpha q_{k+1}, \frac{q_{k+1}-q_k}{h})\; ,
\end{eqnarray*}
with $1\leq k\leq N-1$ and initial condition satisfying 
\begin{eqnarray*}
\tilde{\varphi}^a&(&q_0, q_1, h)=\Psi^a(q_0, -\frac{\partial S^h_{\alpha}}{\partial q_0}(q_0, q_1)\\
&&+ 
(1-\alpha) h \widetilde{\Lambda}((1-\alpha)q_0+\alpha q_1, \frac{q_1-q_0}{h}))=0\; .
\end{eqnarray*}

\begin{example}
{\rm {\bf Nonholonomic particle}.

Consider the Lagrangian $L: T\R^3\rightarrow \R$
\[
L=\frac{1}{2}(\dot{x}^2+\dot{y}^2+\dot{z}^2)-(x^2+y^2)\; ,
\]
 subject to the constraint $\phi=\dot{z}-y\dot{x}=0$.
Taking $\alpha=1/2$ in (\ref{lol}) we obtain a geometric integrator for the continuous nonholonomic problem.
The first figure compares the method introduced here to the
traditional Runge-Kutta method of fourth order, showing an
improvement in several orders of magnitude. Observe that, in this scale,  the value of the energy in each step of our algorithm is practically undistinguishable from the initial value of the energy.
\begin{center}
\includegraphics[width=5.5cm]{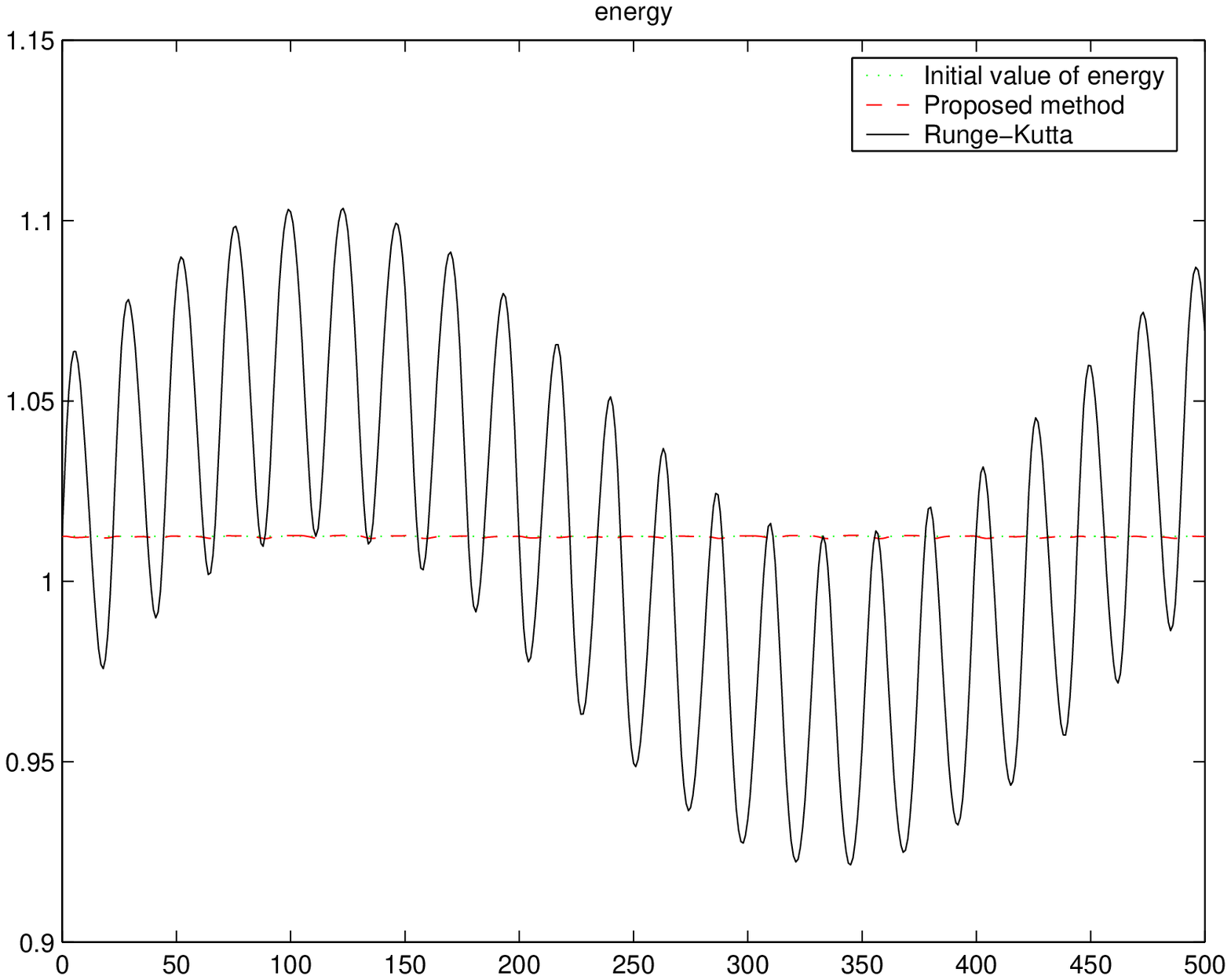}
\end{center}
The second figure is a comparison between our method and the one
proposed in [CorMar:01]. A similar behaviour is observed. 
Nevertheless, a slightly better behaviour can also be appreciated,
where the proposed algorithm shows on average a better
preservation of the original energy.
\begin{center}
\includegraphics[width=5.5cm]{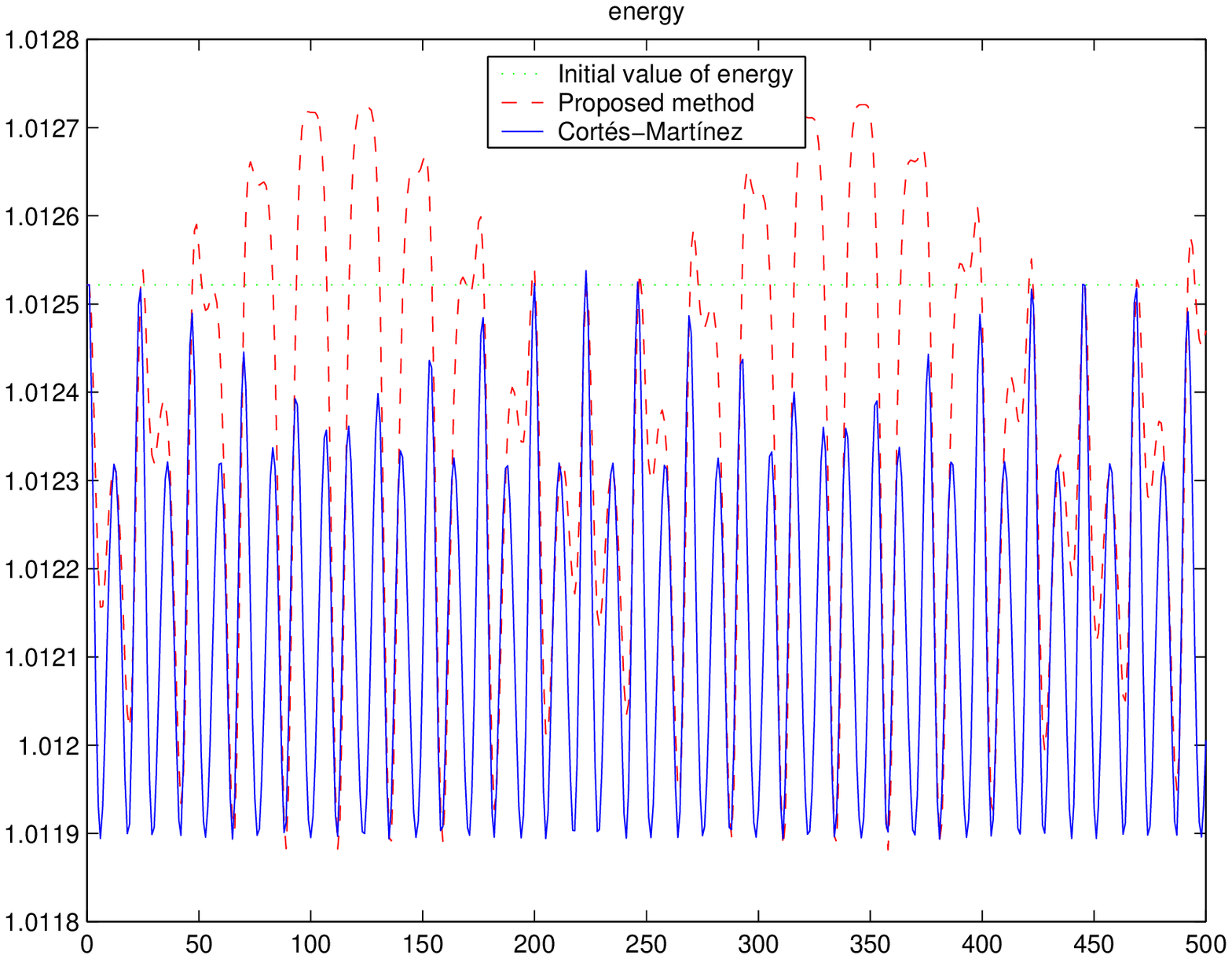}
\end{center}
}
\end{example}

\section{\sc Optimal control theory}

\subsection{Geometric formulation of optimal control problems}

A general optimal control problem consists of a set of differential equations
\begin{equation}\label{general}
\dot{q}^i=\Gamma^i(q(t),u(t)) \,,\; 1 \le i \le n \, ,
\end{equation}
where $q^i$ denote the states and  $u$ the control variables, and 
a cost function $L(q,u)$. Given some boundary conditions (usually $q_0=q(t_0)$ and $q_F=q(t_f)$) the 
aim is to find a $C^2$-piecewise smooth curve $c(t)=(q(t),u(t))$, satisfying the control equations 
(\ref{general}) and minimizing the functional
\begin{equation}\label{general1}
{\mathcal J}(c) = \int^{t_f}_{t_0} L(q(t),u(t))dt \, .
\end{equation}
In a global description, one assumes a fiber bundle structure $\pi: U 
\longrightarrow Q$, where $Q$ is the configuration manifold with local 
coordinates $q^i$ and $U$ is the bundle of controls, with local 
coordinates $(q^i,u^a)$, $1\leq i\leq n$, $1\leq a\leq m$.

The ordinary differential equations (\ref{general}) on $Q$ depending on 
the parameters $u$ can be seen as a vector field $\Gamma$ along the 
projection map $\pi$, that is, $\Gamma$ is a smooth map 
$\Gamma: U \longrightarrow TB$ such that the diagram
\begin{figure}[h]
\centering
\setlength{\unitlength}{1cm}
\begin{picture}(4,2.5)(-0.7,0)
\put(0,2){\makebox(0,0)[r]{$U$}}
\put(4,2){\makebox(0,0)[l]{$TQ$}}
\put(2,0){\makebox(0,0)[c]{$Q$}}
\put(0.2,2){\vector(1,0){3.6}}
\put(2,2.3){\makebox(0,0)[r]{$\Gamma$}}
\put(0.2,1.8){\vector(1,-1){1.6}}
\put(3.8,1.8){\vector(-1,-1){1.6}}
\put(0.8,0.8){\makebox(0,0)[r]{$\pi$}}
\put(3.2,0.8){\makebox(0,0)[l]{$\tau_Q$}}
\end{picture}
\centering
\end{figure}

\noindent is commutative. This vector field is locally written as 
$\Gamma=\displaystyle{\Gamma^i(q,u) \frac{\partial}{\partial q^i}}$.

The solutions of such problem are provided by Pontryaguin's maximum principle. 
If we construct the Hamiltonian function
\begin{equation}\label{hamiltonian}
H(q, p, u)=L(q, u)+p_i \Gamma^i(q, u)
\end{equation}
where $p_i$, $1\leq i\leq n$, are now considered as Lagrange's multipliers, then a curve $\gamma: \R\rightarrow U$, $\gamma(t)=(q(t), u(t))$ is an optimal trajectory  if there exists functions  $p_i(t)$, $1\leq i\leq n$  such that they are  solutions of the Hamilton equations:
\begin{equation}\label{eqH}
\left\{\begin{array}{l}
\dot{q}^i(t)=\displaystyle{\frac{\partial H}{\partial p_i}(q(t), p(t), u(t))}\\
\dot{p}_i(t)=\displaystyle{\frac{\partial H}{\partial q^i}(q(t), p(t), u(t))}\\
\end{array}\right.
\end{equation}
and 
\begin{equation}\label{H}
H(q(t), p(t), u(t))=\underset{v}{\hbox{min}}\; H(q(t), p(t), v), \end{equation}
with $t\in [t_0, t_f]$.
This last condition is usually replaced by 
\begin{equation}\label{constraint}
\frac{\partial H}{\partial u^a}=0, \quad 1\leq a\leq m
\end{equation}
when we are looking for extremal trajectories.

It is well known that the Pontryaguin's necessary conditions for extremality have a geometric interpretation in terms of presymplectic hamiltonian system. 
The total space of the system will be $T^*Q \times_Q U$. 
Let $\omega_Q$ be the canonical symplectic form on $T^*Q$ and consider the canonical projection $\hbox{pr}_1: T^*Q \times_Q U\longrightarrow T^*Q$.
Denote by $\omega=\hbox{pr}^*_1\omega_Q$ the induced closed 2-form on $T^*Q \times_Q U$. 
The 2-form $\omega$ is degenerate and its characteristic distribution is locally spanned by $\partial/\partial u^a$, $1\leq a\leq m$.
Define the Pontryaguin's hamiltonian function $H: T^*Q \times_Q U\longrightarrow \R$ as follows
$H(\alpha_q, u_q)=L(u_q)+\alpha_q(\Gamma(u_q))
$
where $\alpha_q\in T^*_q Q$ and $u_q\in \hbox{pr}^{-1}(q)$.  Obviously, the coordinate expression of $H$ is (\ref{hamiltonian}).

Equations (\ref{eqH})  (\ref{H}) and (\ref{constraint}) are intrinsically written as 
\begin{equation}\label{qqq}
i_X\omega=dH
\end{equation}
Applying the Dirac-Bergmann-Gotay-Nester algorithm to the presymplectic system $(T^*Q \times_Q U, \Omega, H)$ we obtain that equations (\ref{constraint}) correspond to the primary constraints for the presymplectic system: 
$\phi^a=\frac{\partial H}{\partial u^a}=0$.
The equations have solution along the first constraint submanifold $P_0$ determined by the vanishing of the primary constraints. On the points of $P_0$ there is  at least a pointwise solution   of Equation (\ref{qqq}) but such solutions are not, in general, tangent to $P_0$. These points must be removed leaving a subset $P_1\subset P_0$ (it is assumed tan $P_1$ is also  a submanifold).  Then, we have to restrict $P_1$ to a submanifold where  the solutions of (\ref{qqq}) are tangent to $P_1$. Proceeding further, we obtain a sequence of submanifolds
\[
\cdots\hookrightarrow P_k\hookrightarrow\cdots\hookrightarrow P_2\hookrightarrow P_1\hookrightarrow
P_0\hookrightarrow T^*Q\times_Q U
\]
If this algorithm stabilizes, i.e. there exists a positive integer $k\in \ene$ such that $P_k=P_{k+1}$ and $\dim P_k\not= 0$, then we will obtain an stable submanifold $P_f=P_k$, on which a vector field exists such that
\begin{equation}\label{poi}
\left( i_X \omega=dH\right)_{|P_f}
\end{equation}
The constraints determining $P_f$ are known in the control literature as higher order conditions for optimality.
Therefore, a necessary condition for optimality of the curve  $\gamma: \R\rightarrow U$, $\gamma(t)=(q(t), u(t))$  will be the existence of a lift $\tilde{\gamma}$ of $\gamma$ to $P_f$ such that $\tilde{\gamma}$ will be an integral curve of a solution  of Equations (\ref{poi}). 

In the regular case, the final constraint algorithm is $P_0$ (that is, $P_0=P_f$) and all the constraints are second class following the  classical classification of Dirac. In such case $(P_0, \omega_0)$ is a symplectic manifold,  where $\Omega_0$ denotes the restriction of the presymplectic 2-form to the constraint submanifold $P_0$. Locally, the symplecticity of $(P_0, \omega_0)$ is equivalent to the regularity of the matrix
$\displaystyle{\left(
\frac{\partial^2 H}{\partial u^a\partial u^b}
\right)_{1\leq a,b\leq m}}$.
The dynamical equations for the optimal control problem will be 
\begin{equation}\label{aqq}
i_{X_{P_0}}\omega_0=dH_{|P_0}
\end{equation}
Taking coordinates $(q^i, p_i)$ on $P_0$, then the dynamical equations are: 
\begin{equation}\label{eqH1}
\left\{\begin{array}{l} 
\dot{q}^i(t)=\displaystyle{\frac{\partial H_{|P_0}}{\partial p_i}(q(t), p(t))}\\
\dot{p}_i(t)=\displaystyle{\frac{\partial H_{|P_0}}{\partial q^i}(q(t), p(t))}\\
\end{array}\right.
\end{equation}
 where  we have substituted in (\ref{eqH}) the control variables $u^a$ by its value $\bar{u}^a=f^a(q,p)$ applying the implicit function theorem to the primary constraints $\phi^a=0$. 
In such  case, there exists a unique solution $X_{P_0}$ of Equation (\ref{aqq})  and its flow preserves the symplectic 2-form $\omega_0$, i.e. it is a canonical transformation. 

\subsection{Generating functions of the second kind}

Let $({\cal M}, \Omega)$ be an exact  symplectic manifold ($\Omega$ is symplectic and exact, $\Omega=-d\Theta$)  and suppose that $F: {\cal M}\rightarrow {\cal M}$ is a transformation from ${\cal M}$ to itself and $\hbox{Graph}(F)$ the graph of $F$, $\hbox{Graph}(F)\subset {\cal M}\times {\cal M}$.
Denote by $\pi_i: {\cal M}\times {\cal M}\rightarrow {\cal M}$, $i=1,2$ the canonical projections and  the forms:
\begin{eqnarray*}
\bar{\Theta}&=&\pi_2^*\Theta-\pi_1^* \Theta\\
\bar{\Omega}&=&\pi_2^*\Omega-\pi_1^* \Omega=-d\bar{\Theta}\\
\end{eqnarray*}
Denote by $i_F: \hbox{Graph}(F)\hookrightarrow {\cal M}\times {\cal M}$ the inclusion map. Then, $F$ is a canonical transformation if and only if $i_F^*\bar{\Omega}=0$, that is, if $\hbox{Graph}(F)$ is a lagrangian submanifold of $({\cal M}\times {\cal M}, \bar{\Omega})$.
In such a case, $i_F^*\bar{\Omega}=-d i_F^*\bar\Theta=0$ and, at least locally, there exists a function $S: \hbox{Graph}\, F \rightarrow \R$ such that \begin{equation}\label{eee}
i_F^*\bar\Theta=dS
\end{equation}
Taking  $(q^i,p_i)$ as natural coordinates in $\hbox{Graph}(F)$ and $(q^i, p_i, \mathbf{q}^i, \mathbf{p}_i)$ the coordinates in ${\cal M}\times {\cal M}$, then, along $\hbox{Graph}(F)$, $\mathbf{q}^i=\mathbf{q}^i(q,p)$ and $\mathbf{p}_i=\mathbf{p}_i(q,p)$ and
$
\mathbf{p}_i\,d\mathbf{q}^i-p_i\,dq^i=dS(q,p). 
$
Suppose that $(q^i, \mathbf{p}_i)$ are independent local coordinates on $\hbox{Graph}(F)$ (see [Arn:78]); i.e. 
$S=S(q,\mathbf{p})$
Since
\[
\mathbf{p}_i\,d\mathbf{q}^i-p_i\, dq^i=-\mathbf{q}^i\, d\mathbf{p}_i + d(\mathbf{q}^i \mathbf{p}_i)-p_i\, dq^i=dS,\]
if we define $
S_2(q,\mathbf{p})=\mathbf{q}^i \mathbf{p}_i-S(q,\mathbf{p})$,
where $\mathbf{p}$ is expressed in terms of $p$ and $\mathbf{q}$, 
then
$
\mathbf{q}^i\, d\mathbf{p}_i+p_i dq^i=dS_2(q, \mathbf{p})
$

 \begin{definition}
The function $S_2(q,\mathbf{p})$ will be called a {\bf generating function of the second kind} of the canonical transformation $F$.
\end{definition}

Now, suppose that  $({\cal M}, \Omega, H)$ is a hamiltonian system and $X_H$ its hamiltonian vector field, say  $i_{X_H}\Omega=dH$.
Denote by $F_h: {\cal M}\rightarrow {\cal M}$ its flow.  
\begin{theorem}
Let a function $S_2^{Nh}$ be defined by
\[
S_2^{Nh}(q_0, p_{Nh})=\sum_{k=0}^{N-1}(S^h_2(q_k, p_{k+1})-q_{k+1}p_{k+1})
\]
where $q_k$, $1\leq k\leq N$, and $p_k$, $0\leq k\leq N-1$, are stationary points of the 
right-hand side, that is
\begin{eqnarray*}
q_{k+1}&=&\frac{\partial S^h_2}{\partial p}(q_{k}, p_{k+1}), \quad 0\leq k\leq N-1\\ 
p_k&=&\frac{\partial S^h_2}{\partial q}(q_{k}, p_{k+1}), \quad 0\leq k\leq N-1\\ 
\end{eqnarray*}
then $S^{Nh}_2$ is a generating function of the second kind for $F_{Nh}: {\cal M}\rightarrow {\cal M}$.
\end{theorem}
{\bf Proof:}
It is similar to that of Theorem \ref{Th}.\bull

Finally, we have the following 
\begin{proposition}\label{proposition1}
A generating function of the second kind for $F_h$ is given by
\[
S_2^h(q_0,p_h)=p_hq_h-\int^h_0 \left(p\,dq-H\, dt\right)
\]
where $t\rightarrow (q(t), p(t))$ is an integral curve of the Hamilton equations such that $q(0)=q_0$ and $p(h)=p_h$.
\end{proposition}

\subsection{Generating functions of the second kind and discrete optimal control problems}

From Proposition \ref{proposition1} a generating function of the second kind for the Hamiltonian system 
$(P_0, \Omega_{0}, H_{|P_0})$ which determines the dynamics of the optimal control problem given by (\ref{general}) and (\ref{general1}) is 
\begin{eqnarray}\label{ppp}
&&
S^h_2(q_0, p_h)=p_hq_h\nonumber\\
&&-\int_0^h \left(p(t)\dot{q}(t)-H_{|P_0}(q(t),p(t))\right)\,dt
\end{eqnarray}
where $t\rightarrow (q(t), p(t))$ is an integral curve of the vector field $X_{P_0}$
with $(q(0), p(0))=(q_0, p_0)$ and $(q(h), p(h))=(q_h, p_h)$.

We now turn to the construction of a numerical  integrator for the Hamiltonian system $(P_0, \omega_{0}, H_{|P_0})$ 
by using an approximation of the generating function. The proposed methods  also realize the integration steps by canonical transformations; therefore, they are symplectic integrators.

\begin{example}
{\rm 
Consider, for instance,  the following approximation to $S^h_2$:
\begin{eqnarray*}
&&\tilde{S}^h_2(q_k, p_{k+1})=p_{k+1}q_{k+1}-hp_{k+1}\left(\frac{q_{k+1}-q_k}{h}\right)\\
&&+h\tilde{L}_d(q_k, p_{k+1})+hp_{k+1}\tilde{\Gamma}_d(q_k, p_{k+1})
\end{eqnarray*}
where $\tilde{L}_d$ and $\tilde{\Gamma}_d$ are adequate approximations to $L_{|P_0}$ and $\Gamma_{|P_0}$, respectively.

Denote by $\tilde{f}(q_k, p_{k+1})$ the function
$\tilde{f}(q_k, p_{k+1})=h\Gamma_d(q_k, p_{k+1})+q_{k}$. Since
$\displaystyle{\frac{q_{k+1}-q_k}{h}=\tilde{\Gamma}_d(q_k, p_{k+1})
}$ then, 
\[
\tilde{S}^h_2(q_k, p_{k+1})=\tilde{L}_d(q_k, p_{k+1}) +p_{k+1}\tilde{f}(q_k, p_{k+1})
\]
and hence the equations 
\begin{equation}\label{aq4}
\left\{
\begin{array}{l}
\displaystyle{p_k= \frac{\partial \tilde{S}^h_d}{\partial q}(q_k, p_{k+1})}\\
\displaystyle{q_{k+1}= \frac{\partial \tilde{S}^h_d}{\partial p}(q_k, p_{k+1})}\\
\end{array}
\right.
\end{equation}
are exactly the discrete equations corresponding to the classical discrete optimal control problem (see [Lew:86]), determined by the control equations:
$
q^i_{k+1}=\tilde{f}^i(q_k, u_k),\quad   \hbox{($(q_0)$ given)}
$
and with associate perfomance index:
$
J=\sum_{k=0}^{N-1} \tilde{L}_d(q_k, u_k)
$
Observe that this discrete optimal control problem is symplectic in the sense explained in the  subsection above. 
}
\end{example}



\end{document}